\documentclass[8pt,a4paper]{article}
\usepackage{a4wide}
\usepackage{inputenc}
\usepackage{amsmath}
\usepackage{amssymb}
\usepackage{float}
\usepackage{xspace}
\usepackage{multicol}
\usepackage{color}
\usepackage{url}

\definecolor{oxygenorange}{rgb}{0.97,0.5,0.04}
\definecolor{oxygengray}{rgb}{0.35,0.35,0.35}
\definecolor{oxygenlightgray}{rgb}{0.93,0.93,0.93}
\definecolor{oxygenblue}{rgb}{0.14, 0.43, 0.69}

\newcommand{\ord}[1]{\ensuremath{\mathcal{O}\!\left(#1\right)}}
\newcommand{\field}[1]{\mathbb{#1}}
\newcommand{\GFZ}{\field{F}_2\xspace}
\newcommand{\Magma}{\textsc{Magma}\xspace}
\newcommand{\PolyBoRi}{\textsc{PolyBoRi}\xspace}
\newcommand{\Sage}{Sage\xspace}
\newcommand{\GAP}{GAP\xspace}
\newcommand{\xBG}{\textsf{x86\_64}\xspace}
\newcommand{\Opteron}{\textsf{Opteron}\xspace}
\newcommand{\CTD}{\textsf{Core 2 Duo}\xspace}

\newfloat{example}{htbp}{loa}
\floatname{example}{Example}

\floatstyle{ruled}
\newfloat{algorithm}{htbp}{loa}
\floatname{algorithm}{Algorithm}

\definecolor{darkgray}{rgb}{0.4,0.4,0.4} 
\definecolor{lightgray}{rgb}{0.7,0.7,0.7} 

\title{Efficient Multiplication of Dense Matrices over GF(2)}
\author{MARTIN ALBRECHT\\Information Security Group, Royal Holloway, University
of London\\ GREGORY BARD\\Department of Mathematics, Fordham University \and
WILLIAM HART\\Department of Mathematics, University of Warwick}

\begin{document}

\begin{abstract}
We describe an efficient implementation of a hierarchy of
algorithms for multiplication of dense matrices over the field with two elements
($\GFZ$). In particular we present our implementation -- in the M4RI
library -- of Strassen-Winograd matrix multiplication and the ``Method of the
Four Russians'' multiplication (M4RM) and compare it against
other available implementations. Good performance is demonstrated on
on AMD's \Opteron and particulary good performance on Intel's \CTD. The
open-source M4RI library is available stand-alone as well as part of the
\Sage mathematics software.

In machine terms, addition in $\GFZ$ is logical-XOR, and multiplication
is logical-AND, thus a machine word of 64-bits allows one to operate on 64
elements of $\GFZ$ in parallel: at most one CPU cycle for 64 parallel additions
or multiplications. As such, element-wise operations over $\GFZ$ are relatively
cheap. In fact, in this paper, we conclude that the actual bottlenecks are
memory reads and writes and issues of data locality. We present our
empirical findings in relation to minimizing these and give an analysis thereof.
\end{abstract}

\maketitle

\section{Introduction}
We describe an efficient implementation of a hierarchy of
algorithms for multiplication of dense matrices over the field with two elements
($\GFZ$). Ma\-trix-matrix multiplication is an important primitive in
computational linear algebra and as such the fundamental algorithms we implement
have been well-known for some time. Therefore this paper focuses on the numerous
techniques employed for the special case of $\GFZ$ in the M4RI library
(\url{http://m4ri.sagemath.org}) and the benefits so derived.

We note that even for problems that do not reduce to matrix-matrix
multiplication many of the techniques presented in this paper are still
applicable. For instance, Gaussian Elimination can be achieved via the ``Method
of the Four Russians'' Inversion (M4RI)(cf. \cite[Ch. 5]{bard-phd} and
\cite{bard-m4ri}) and borrows ideas from the ``Method of the Four Russians''
Multiplication (M4RM)~\cite{ADKF70}, \cite{AHU74} which we present here.

The M4RI library implements dense linear algebra over $\GFZ$ and is used by
\Sage \cite{sage} and \PolyBoRi~\cite{polybori}.

Our optimization efforts focus on 64 bit x86 architectures (\xBG),
specifically the Intel \CTD and the AMD \Opteron. Thus, we assume in
this paper that each native CPU word has 64-bits: $w_s=64$. However it should be
noted that our code also runs on 32-bit CPUs and on non-x86 CPUs such as the
PowerPC.

In machine terms, addition in $\GFZ$ is logical-XOR, and multiplication
is logical-AND, thus a machine word of 64-bits allows one to operate on 64
elements of $\GFZ$ in parallel: at most one CPU cycle for 64 parallel additions
or multiplications. As such, element-wise operations over $\GFZ$ are relatively
cheap. In fact, in this paper, we conclude that the actual bottlenecks are
memory reads and writes and issues of data locality. We present our
empirical findings in relation to minimizing these and give an analysis thereof.

The second author proposed, in \cite{Bard2006} and \cite[Ch. 5]{bard-phd}, to
count memory accesses rather than arithmetic operations to estimate the
complexity of such algorithms and the empirical results of this
paper lend further support to this model. However, this model is a
simplification as memory access is not uniform, i.e. an algorithm which randomly
accesses memory will perform much worse than an algorithm with better spatial
and temporal locality. While these differences only affect the constant of a
complexity estimation, in practice they make a very significant difference, as
our results will demonstrate. 

The paper is structured as follows. We proceed from basic arithmetic
(Section~\ref{sec:basic}) via the classical cubic multiplication algorithm
(Section~\ref{sec:cubic}), through a detailed discussion of the ``Method of the
Four Russians'' (Section~\ref{sec:m4rm}) to the Strassen-Winograd algorithm
(Section~\ref{sec:strassen}).
We start by introducing our basic data structures and conclude by presenting
timing experiments to show the validity of our approach
(Section~\ref{sec:benchmark}).
Note, that all timings in this paper time Strassen-Winograd multiplication (cf.
Section~\ref{sec:strassen}) but with different base cases.

\section{Basic Arithmetic}
\label{sec:basic}
\subsection{Our Matrix Data Structure}
We use a ``flat row-major representation'' for our matrices. Thus
64 consecutive entries in one row are packed into one
machine word. Consequently, bulk operations on whole rows are
considerably cheaper than on whole columns and addressing a single column is
more expensive than addressing a single row. Additionally, we maintain an array
-- called \texttt{rowswap} -- containing the address in memory of the first word
for each row in the matrix. To represent in-place submatrices (i.e. without
copying out the data) we also use this \texttt{rowswap} array. We call these
in-place submatrices ``matrix windows'' and they consist of addresses of the
first word of each row and the number of columns each row contains. This
approach is limited to ``matrix windows'' which start and end at full word
borders but this is sufficient for our application. The advantages and
disadvantages of the ``flat row-major'' data structure are, for instance,
analyzed in \cite{fflas}.

\subsection{Row Additions}
Since this basic operation -- addition of two rows -- is at the heart of every
algorithm in this paper we should briefly mention the SSE2 instruction set
\cite{optcpp} which is available on modern \xBG architectures. This instruction
set offers an XOR operation for 128-bit wide registers, allowing one to handle
two 64-bit machine words in one instruction. The use of these instructions does
provide a considerable speed improvement on Intel
CPUs. Table~\ref{tab:m4rm-sse2-c2d} shows that up to a 25\% improvement is
possible when enabling SSE2 instructions. However, in our experiments
performance declined on \Opteron CPUs when using SSE2 instructions. The authors
were unable to identify a cause of this phenomenon.

\begin{table}[htbp]
\begin{center}
\begin{tabular}{|c|r|r|}
\hline
Matrix Dimensions      &  Using 64-bit & Using 128-bit (SSE2)\\
\hline
$10,000 \times 10,000$ &         1.981 & 1.504\\ 
$16,384 \times 16,384$ &         7,906 &  6.074\\ 
$20,000 \times 20,000$ &        14.076 & 10.721\\ 
$32,000 \times 32,000$ &        56.931 & 43.197\\ 
\hline
\end{tabular}
\caption{Strassen-Winograd multiplication on 64-bit Linux, 2.33Ghz
\CTD}
\label{tab:m4rm-sse2-c2d}
\end{center}
\end{table}

\subsection{Cubic Multiplication}
\label{sec:cubic}
The simplest multiplication operation involving matrices is a matrix-vector
product which can easily be extended to classical cubic matrix-matrix
multiplication.
To compute the matrix-vector product $Ab$ we have to compute the dot
product of each row $i$ of $A$ and the vector $b$. If the vector $b$ is
stored as a row rather than a column, this calculation becomes equivalent to
word-wise logical-AND and accumulation of the result in a word $p$ via
logical-XOR. Finally, the parity of $p$ needs to be computed. 
However, as there is no native parity instruction in the \xBG instruction set
this last step is quite expensive compared to the rest of the routine. To
account for this, 64 parity bits can be computed in
parallel~\cite[Ch. 5]{hackersdelight}. To extend this matrix-vector
multiplication to matrix-matrix multiplication $B$ must be stored transposed.

\section{The Method of the Four Russians}
\label{sec:m4rm}
The ``Method of the Four Russians'' matrix
multiplication algorithm can be derived from the original algorithm published
by Arlazarov, Dinic, Kronrod, and Faradzev \cite{ADKF70}, but does not directly
appear there. It has appeared in books including \cite[Ch. 6]{AHU74}.

Consider a product of two matrices $C = AB$ where $A$ is an $m \times l$ matrix
and $B$ is an $l \times n$ matrix, yielding an $m \times n$ for $C$. $A$ can be
divided into $l/k$ vertical ``stripes'' $A_0 \dots
A_{(l-1)/k}$ of $k$ columns each, and $B$ into $l/k$ horizontal stripes $B_0
\dots B_{(l-1)/k}$ of $k$ rows each. (For simplicity assume $k$ divides $l$).
The product of two stripes, $A_i B_i$ requires an $m \times l/k$ by $l/k \times
n$
matrix multiplication, and yields an $m \times n$ matrix $C_i$. The sum of all
$k$ of these $C_i$ equals $C$.
\[
C = AB = \sum_0^{(l-1)/k} A_iB_i. 
\]

\emph{Example:} Consider $k=1$ and
\[
A = \left(\begin{array}{rr}
a_{0} & a_{1} \\
a_{2} & a_{3}
\end{array}\right), B = \left(\begin{array}{rr}
b_{0} & b_{1} \\
b_{2} & b_{3}
\end{array}\right).
\]
Then \[
A_0 = \left(\begin{array}{r}
a_{0} \\
a_{2}
\end{array}\right), A_1 = 
\left(\begin{array}{r}
a_{1} \\
a_{3}
\end{array}\right), B_0 = \left(\begin{array}{rr}
b_{0} & b_{1}
\end{array}\right), \textnormal{ and } B_1 = 
\left(\begin{array}{rr}
b_{2} & b_{3}
\end{array}\right)
\] and consequently
\[
A_0 B_0 = \left(\begin{array}{rr}
a_{0} b_{0} & a_{0} b_{1} \\
a_{2} b_{0} & a_{2} b_{1}
\end{array}\right) \textnormal{ and } A_1 B_1 = \left(\begin{array}{rr}
a_{1} b_{2} & a_{1} b_{3} \\
a_{3} b_{2} & a_{3} b_{3}
\end{array}\right).
\] Finally, we have
\[
C = AB = A_0B_0 + A_1B_1 = \left(\begin{array}{rr}
a_{0} b_{0} + a_{1} b_{2} & a_{0} b_{1} + a_{1} b_{3} \\
a_{2} b_{0} + a_{3} b_{2} & a_{2} b_{1} + a_{3} b_{3}
\end{array}\right).
\]
\vspace{0.2cm}

The principal benefit of multiplying in narrow stripes is that the bits across
each row of a stripe of $A$ determine which linear combination of rows of $B$
will contribute to the product, e.g. in the above example $a_0, \dots , a_3$
dictate which linear combination of $b_0$, $b_2$ and $b_1, b_3$ must be
written to the rows of $C$. However, if the stripe is
relatively narrow as in this example, there is only a small 
number of binary values each row of the stripe can take, and thus only
a small number of possible linear combinations of the rows of $B$ that
will be ``selected''. If we precompute all possible linear
combinations of rows of $B$ that could be selected we can create a
lookup table into which the rows of the stripes of $A$ can index.

Returning to our example, if $a_0 = a_2$ and $a_1 = a_3$ then the same linear
combination would be written to the first and the second row of $C$.
Precomputation of all $2^4-1$ non-zero linear combinations, ($1\cdot b_0+
0\cdot b_1,0\cdot b_0+ 1\cdot b_1,1\cdot b_0+ 1\cdot b_1$), ensures that the
repeated linear combination has only been computed once. In our
trivial example this is not a saving, but for much larger matrices
reuse of the precomputed combinations gives a saving. Precomputing a table in
this fashion is also called ``greasing''.

The technique just described gives rise to Algorithm~\ref{alg:m4rm}.
\begin{algorithm}
\caption{M4RM}
\begin{tabular}{l}
function AddRowFromTable(C, $r_1$, T, $r_2$) begin\\
\hspace{0.33in}for $0 \le i <$ NumberOfColumns(C) do begin\\
\hspace{0.66in}$C_{r_1,i} \leftarrow C_{r_1,i} + T_{r_2,i}$\\
\hspace{0.33in}end\\
end\\
\\
function ReadBits(A, r, c, k) begin\\
\hspace{0.33in}return $A_{r,c}*2^{k-1} + A_{r,c+1}*2^{k-2} + A_{r,c+2}*2^{k-3} + \cdots + A_{r,c+k-1}*2^0$\\
end\\
\\
function MethodFourRussiansMultiplication(A, B, k) do begin\\
\hspace{0.33in}$m \leftarrow$ NumberOfRows(A)\\
\hspace{0.33in}$\ell \leftarrow$ NumberOfColumns(A)\\ 
\hspace{0.33in}$n \leftarrow$ NumberOfColumns(B)\\
\hspace{0.33in}$C \leftarrow$ GenerateZeroMatrix(m, n)\\
\\
\hspace{0.33in}for $0 \le i < \mbox{floor}(\ell/k)$ do begin\\
\hspace{0.66in}//create table of $2^k-1$ linear combinations\\
\hspace{0.66in}$T \leftarrow$ MakeTable(B, i*k, 0, k)\\
\hspace{0.66in}for $0 \le j < m$ do begin\\
\hspace{0.99in}//read index for table T\\
\hspace{0.99in}id $\leftarrow$ ReadBits(A, j, k*i, k)\\
\hspace{0.99in}//add appropriate row from table T\\
\hspace{0.99in}AddRowFromTable(C, j, T, id)\\
\hspace{0.66in}end\\
\hspace{0.33in}end\\
\hspace{0.33in}return C\\
end
\end{tabular}
\label{alg:m4rm}
\end{algorithm}
In Algorithm~\ref{alg:m4rm} the subroutine \texttt{ReadBits(A, r, sc, k)}
reads $k$ bits from row $r$
starting at column $sc$ and returns the bit string
interpreted as an integer and \texttt{AddRowFromTable(C, r,
T, x)} adds the row $x$ from $T$ to the row $j$ of $C$. The subroutine
\texttt{MakeTable(B, r, c, k)} in Algorithm~\ref{alg:m4rm} constructs a table
$T$ of all $2^k-1$ non-zero linear combinations of the rows of $B$ starting in
row $r$ and column $c$. For this calculation Gray codes are used.

\subsection{Gray Codes}
The Gray code~\cite{graycode}, named after Frank Gray and also known as
reflected binary code, is a numbering system where two consecutive values
differ in only one digit. Examples of Gray codes for two, three and four bits
are given in
Figure~\ref{fig:graycodes}.

\begin{figure}[htbp]
\label{fig:graycodes}
\begin{center}
\begin{multicols}{2}
\begin{tabular}{cc}
0 & 0\\
0 & 1\\
1 & 1\\
1 & 0\\
\end{tabular}\\
2-bit Gray Code\\
\vspace{2em}
\begin{tabular}{ccc}
0 & 0 & 0\\
0 & 0 & 1\\
0 & 1 & 1\\
0 & 1 & 0\\
1 & 1 & 0\\
1 & 1 & 1\\
1 & 0 & 1\\
1 & 0 & 0\\
\end{tabular}\\
3-bit Gray Code\\

\begin{tabular}{cccc}
0 & 0 & 0 & 0\\
0 & 0 & 0 & 1\\
0 & 0 & 1 & 1\\
0 & 0 & 1 & 0\\
0 & 1 & 1 & 0\\
0 & 1 & 1 & 1\\
0 & 1 & 0 & 1\\
0 & 1 & 0 & 0\\
1 & 1 & 0 & 0\\
1 & 1 & 0 & 1\\
1 & 1 & 1 & 1\\
1 & 1 & 1 & 0\\
1 & 0 & 1 & 0\\
1 & 0 & 1 & 1\\
1 & 0 & 0 & 1\\
1 & 0 & 0 & 0\\
\end{tabular}\\
4-bit Gray Code\\

\end{multicols}
\end{center}

\caption{Gray Codes}
\end{figure}

Gray code tables for $n$-bits can be computed efficiently from $n-1$-bit Gray
code tables by prepending each entry of the $n-1$-bit Gray code table
with $0$. Then the order of the entries is reversed and a $1$ is prepended
to each entry. These two half-tables are then concatenated. These tables can
then be used to construct all $2^k-1$ non-zero linear combinations of $k$ rows
where each new entry in the table costs one row addition as its
index differs in exactly one bit from that of the preceding row. Thus computing
all $2^k-1$ non-zero linear combinations of $k$ rows can be done in $2^k-1$ row
additions, rather than $(k/2-1)2^k-1$ as would be expected if each vector were
to be tabulated separately.

From the complexity analysis in \cite{Bard2006} it seems one should always 
choose the parameter $k = \lfloor \log_2 n \rceil$ for an $n \times n$ matrix.
However, in practice this 
is not the case. First, experimental evidence indicates \cite{bard-phd} that 
$0.75 \times \log_2 n$ seems to be a better choice. Also, for cache 
efficiency it makes sense to split the input matrices into blocks such that 
these blocks fit into L2 cache (see below). If that technique is employed 
then the block sizes dictate $k$ and not the total dimensions of the input 
matrices. 
Thus, a much smaller $k$ than $\log_2 n$ is found to be
optimal, in practice (see below); restraining $k$ in this 
way actually improves performance.

We pre-compute the Gray Code tables up to
size 16. For matrices of dimension $> 20$ million rows and columns, this is not
enough. But, such a dense matrix would have nearly half a quadrillion entries,
and this is currently beyond the capabilities of existing computational
hardware. Also, for these dimensions the Strassen-Winograd algorithm should be
used.

\subsection{A Cache Friendly Version}
Note that the M4RM algorithm creates a table for each stripe of $B$ and then
iterates over all rows of $C$ and $A$ in the inner loop. If the matrices $C$
and $A$ are bigger than L2 cache then this means that for each single row
addition a new row needs to be loaded from RAM. This row will evict an older row
from L2. However, as this row is used only once per iteration of all rows of
$A$ and $C$ we cannot take advantage of the fact that it is now in L2 cache.
Thus if the matrices $A$ and $C$ do not fit into L2 cache then the algorithm
does not utilize this faster memory.

Thus, it is advantageous to re-arrange the algorithm in such a way that it
iterates over the upper part of $A$ completely with all tables for $B$
before going on to the next part. This gives rise to
Algorithm~\ref{alg:m4rm_cf}, a cache friendly version of the M4RM algorithm. For
simplicity we assume that $m,l,n$ are all multiples of
some fixed block size in the presentation of Algorithm~\ref{alg:m4rm_cf}.
\begin{algorithm}
\caption{Cache Friendly M4RM}
\begin{tabular}{l}
function MethodOfFourRussiansCacheFriendlyMultipication(A, B, k)\\
\hspace{0.33in}$m \leftarrow$ NumberOfRows(A)\\
\hspace{0.33in}$\ell \leftarrow$ NumberOfColumns(A)\\
\hspace{0.33in}$n \leftarrow$ NumberOfColumns(B)\\
\hspace{0.33in}$C \leftarrow$ GenerateZeroMatrix(m, n)\\
\\
\hspace{0.33in}for $0 \le$ start $<$ m/BlockSize do begin\\
\hspace{0.66in}for  $0<= i < \ell/k$ do begin\\
\hspace{0.99in}$T \leftarrow$ MakeTable(B, i*k, 0, k)\\
\hspace{0.99in}for 0  $\le s <$ BlockSize do begin\\
\hspace{1.33in}$j \leftarrow$ start * BlockSize + s\\
\hspace{1.33in}$x \leftarrow$ ReadBits(A, j, k*i, k)\\
\hspace{1.33in}AddRowFromTable(C, j, T, id)\\
\hspace{0.99in}end\\
\hspace{0.66in}end\\
\hspace{0.33in}end\\
\hspace{0.33in}return C\\
end\\
\end{tabular}
\label{alg:m4rm_cf}
\end{algorithm}
This cache-friendly rearrangement is paid for by the repeated regeneration of
the table $T$. However, compared to the inner loop, this is a cheap operation
and thus is outweighed by the better data locality.
Table~\ref{tab:m4rm-techniques-c2d} shows that this strategy provides
considerable performance improvements.

\subsection{Increasing the Number of Gray Code Tables}
Recall that the actual arithmetic is quite cheap compared to memory 
reads and writes and that the cost of memory accesses greatly
depends on where in memory data is located:
the L1 cache is approximately 50 times faster than main memory. It is thus
advantageous to try to fill all of L1
with Gray code tables. For example consider $n = 10000$, $k=10$ and one Gray
code table. In this situation we work on 10 bits at a time. If we use $k=9$
and
two Gray code tables, we still use the same memory for the tables but can deal
with 18 bits at once. The price we pay is one additional row addition,
which is cheap if the operands are all in cache. To implement this
enhancement the algorithm remains almost unchanged, except that $t$ tables are
generated for $tk$ consecutive rows of $B$, $tk$ values $x$ are read for
consecutive entries in $A$ and $t$ rows from $t$ different tables are added to
the target row of $C$. This gives rise to Algorithm~\ref{alg:m4rm_2t} where we
assume that $tk$ divides $l$ and fix $t=2$.

\begin{algorithm}
\caption{M4RM with Two Gray Code Tables}
\begin{tabular}{l}
function AddTwoRowsFromTable(C, $r_0$, $T$, $r_1$, $TT$, $r_2$) do begin\\
\hspace{0.33in}for $0 <= i <$ NumberOfColumns(C) do begin\\
\hspace{0.66in}$C_{r,i} \leftarrow C_{r,i} + T_{r_1,i} + TT_{r_2,i}$\\
\hspace{0.33in}end\\
end\\
\\
function MethodOfFourRussiansTwoTables(A, B, k) do begin\\
\hspace{0.33in}$m \leftarrow$ NumberOfRows(A)\\
\hspace{0.33in}$\ell \leftarrow$ NumberOfColumns(A)\\ 
\hspace{0.33in}$n \leftarrow$ NumberOfColumns(B)\\
\hspace{0.33in}$C \leftarrow$ GenerateZeroMatrix(m, n)\\
\\
\hspace{0.33in}for $0 \le i < \ell/(2*k)$ do begin\\
\hspace{0.66in}$T \leftarrow$ MakeTable(B, 2*i*k, 0, k)\\
\hspace{0.66in}$TT \leftarrow$ MakeTable(B, 2*i*k + k, 0, k)\\
\hspace{0.66in}for $0 \le j < m$ do begin\\
\hspace{0.99in}$r_1 \leftarrow$ ReadBits(A, j, 2*k*i, k)\\
\hspace{0.99in}$r_2 \leftarrow$ ReadBits(A, j, 2*k*i+k, k)\\
\hspace{0.99in}AddTwoRowsFromTable(C, j, T, $r_1$, TT, $r_2$)\\
\hspace{0.66in}end\\
\hspace{0.33in}end\\
\hspace{0.33in}return C\\
end\\
\end{tabular}
\label{alg:m4rm_2t}
\end{algorithm}

Table~\ref{tab:m4rm-techniques-c2d} shows that increasing the number of tables
is advantageous. Our implementation uses eight Gray code tables,
which appears to be a good default value according to our experiments.

\begin{table}[htbp]
\begin{footnotesize}
\begin{center}
\begin{tabular}{|c|r|r|r|r|}
\hline
 & \multicolumn{4}{c|}{``base cases'' (cf. Section~\ref{sec:tuning})}\\
\hline
Matrix Dimensions & 
Algorithm~\ref{alg:m4rm} &
Algorithm~\ref{alg:m4rm_cf} & 
Algorithm~\ref{alg:m4rm_2t}, $t=2$ &
Algorithm~\ref{alg:m4rm_2t}, $t=8$ \\
\hline
$10,000\times10,000$ &  4.141 &  2.866 &  1.982 &  1.599\\
$16,384\times16,384$ & 16.434 & 12.214 &  7.258 &  6.034\\ 
$20,000\times20,000$ & 29.520 & 20.497 & 14.655 & 11.655\\
$32,000\times32,000$ & 86.153 & 82.446 & 49.768 & 44.999\\
\hline
\end{tabular}
\caption{Strassen-Winograd with different base cases on 64-bit
Linux, 2.33Ghz
\CTD}
\label{tab:m4rm-techniques-c2d}
\end{center}
\end{footnotesize}
\end{table}

\section{Strassen-Winograd Multiplication}
\label{sec:strassen}
In 1969 Volker Strassen \cite{strassen} published an
algorithm which multiplies two block matrices 
\[
 A = \left(\begin{array}{cc}
           A_{00} & A_{01}\\
           A_{10} & A_{11}\\
           \end{array}\right)
 B = \left(\begin{array}{cc}
           B_{00} & B_{01}\\
           B_{10} & B_{11}\\
           \end{array}\right)
\]
with only seven submatrix multiplications and 18 submatrix additions rather
than eight multiplications and eight additions. As matrix multiplication
($\ord{n^\omega}$, $\omega \geq 2$) is considered more expensive than matrix
addition ($\ord{n^2}$) this is an improvement. Later the
algorithm was improved by Winograd to use 15 submatrix
additions only, the result is commonly referred to as Strassen-Winograd
multiplication.
While both algorithms are to a degree less numerically stable than
classical cubic
multiplication over floating point numbers~\cite[Ch. 26.3.2]{Higham} this
problem does not affect matrices over finite fields and thus the improved
complexity of $\ord{n^{\log_2 7}}$ \cite{strassen,bard-phd} is applicable here.

Let $m$, $l$ and $n$ be powers of two. Let $A$ and $B$ be two matrices of
dimension $m \times l$ and $l \times n$ and let $C = A \times B$. Consider the
block decomposition
\[
 \left(\begin{array}{cc}
           C_{00} & C_{01}\\
           C_{10} & C_{11}\\
           \end{array}\right)
  = \left(\begin{array}{cc}
           A_{00} & A_{01}\\
           A_{10} & A_{11}\\
           \end{array}\right)
  \left(\begin{array}{cc}
           B_{00} & B_{01}\\
           B_{10} & B_{11}\\
           \end{array}\right)
\]
where $A_{00}$ and $B_{00}$ have dimensions $m/2 \times l/2$ and $l/2
\times n/2$ respectively. The Strassen-Winograd algorithm, which computes the $m
\times n$ matrix $C = A \times B$, is given in Algorithm~\ref{alg:strassen}.

\begin{algorithm}
\caption{Strassen-Winograd}
\begin{tabular}{l|l}
function StrassenWinograd(A,B) do begin& \\
\hspace{0.00in}$m \leftarrow$ NumberOfRows(A)& \hspace{0.00in}//7 recursive multiplications\\
\hspace{0.00in}$\ell \leftarrow$ NumberOfColumns(A)&  \hspace{0.00in}$P_0 \leftarrow$ Multiply$(A_{NW}, B_{NW})$\\
\hspace{0.00in}$n \leftarrow$ NumberOfColumns(B)& \hspace{0.00in}$P_1 \leftarrow$ Multiply$(A_{NE}, B_{SW})$\\
\hspace{0.00in}$A_{NW} \leftarrow$ SubMatrix$(A_{0,0} \ldots A_{m/2-1, \ell/2-1})$&
\hspace{0.00in}$P_2 \leftarrow$ Multiply$(S_3, B_{SE})$\\
\hspace{0.00in}$A_{NE} \leftarrow$ SubMatrix$(A_{0,l/2} \ldots A_{m/2-1,
\ell-1})$&
\hspace{0.00in}$P_3 \leftarrow$ Multiply$(A_{SE}, T_3)$\\
\hspace{0.00in}$A_{SW} \leftarrow$ SubMatrix$(A_{m/2,0} \ldots A_{m-1,
\ell/2-1})$&
\hspace{0.00in}$P_4 \leftarrow$ Multiply$(S_0, T_0)$\\
\hspace{0.00in}$A_{SE} \leftarrow$ SubMatrix$(A_{m/2,\ell/2} \ldots A_{m-1,
\ell-1})$&
\hspace{0.00in}$P_5 \leftarrow$ Multiply$(S_1, T_1)$\\
&
\hspace{0.00in}$P_6 \leftarrow$ Multiply$(S_2, T_2)$\\
\hspace{0.00in}$B_{NW} \leftarrow$ SubMatrix$(B_{0,0} \ldots B_{\ell/2-1,
n/2-1})$ &
\\
\hspace{0.00in}$B_{NE} \leftarrow$ SubMatrix$(B_{0,n/2} \ldots B_{\ell/2-1,
n-1})$ &
\hspace{0.00in}//7 final additions\\
\hspace{0.00in}$B_{SW} \leftarrow$ SubMatrix$(B_{\ell/2,0} \ldots B_{\ell-1,
n/2-1})$ &
\hspace{0.00in}$U_0 \leftarrow P_0 + P_1$\\ 
\hspace{0.00in}$B_{SE} \leftarrow$ SubMatrix$(B_{\ell/2,n/2} \ldots B_{\ell-1,
n-1})$ &
\hspace{0.00in}$U_1 \leftarrow P_0 + P_5$\\
&
\hspace{0.00in}$U_2 \leftarrow U_1 + P_6$ \\
\hspace{0.00in}//8 additions&
\hspace{0.00in}$U_3 \leftarrow U_1 + P_4$ \\
\hspace{0.00in}$S_0 \leftarrow A_{SW} + A_{SE}$&
\hspace{0.00in}$U_4 \leftarrow U_3 + P_2$ \\ 
\hspace{0.00in}$S_1 \leftarrow S_0 - A_{NW}$&
\hspace{0.00in}$U_5 \leftarrow U_2 - P_3$\\
\hspace{0.00in}$S_2 \leftarrow A_{NW} - A_{SW}$ &
\hspace{0.00in}$U_6 \leftarrow U_2 + P_4$\\ 
\hspace{0.00in}$S_3 \leftarrow A_{NE} - S_1$&
\\
\hspace{0.00in}$T_0 \leftarrow B_{NE} - B_{NW}$&
\hspace{0.00in}$C_N \leftarrow$ Augment$(U_0, U_4)$\\
\hspace{0.00in}$T_1 \leftarrow B_{SE} - T_0$&
\hspace{0.00in}$C_S \leftarrow$ Augment$(U_5, U_6)$\\
\hspace{0.00in}$T_2 \leftarrow B_{SE} - B_{NE}$&
\hspace{0.00in}$C \leftarrow$ Stack$(C_N,C_S)$\\
\hspace{0.00in}$T_3 \leftarrow T_1 - B_{SW}$&
\hspace{0.00in}return C\\
& end\\
\end{tabular}
\label{alg:strassen}
\end{algorithm}

The subroutine \texttt{Augment} in Algorithm~\ref{alg:strassen} takes two $m
\times l$ and $m \times n$ matrices $A$ and $B$ and returns the $m \times (n+l)$
matrix $C = (A\ B)$ and the subroutine \texttt{Stack} takes two $m \times n$ and
$l \times n$ matrices $A$ and $B$ and returns the $(m + l) \times n$ matrix $$C
= \left(\begin{array}{c}A\\B\end{array}\right).$$

At each recursion step the matrix dimensions must be divisible by two which
explains the requirement of them being powers of two. However, in practice the
recursion stops at a given \emph{cutoff} dimension ($c_o$) and switches over to
another multiplication algorithm. In our case, this is the M4RM algorithm. Thus
the requirement can be relaxed to the requirement that for each recursion step
the matrix
dimensions must be divisible by two.

However, this still is not general enough. Additionally, in case of $\GFZ$ the
optimal case is when $m,n,l$ are  64 times powers of 2 to avoid cutting
within words. To deal with odd-dimensional matrices two strategies are known in
the literature~\cite{strassen-implementation}: One can
either increase the matrix dimensions -- this is called ``padding'' -- to the
next ``good'' value and fill the additional entries with zeros, yielding $A^+$
and $B^+$. Then one can compute $C^+ = A^+B^+$ and finally cut out the actual
product matrix $C$ from the bigger matrix $C^+$. A variant of this approach is
to only virtually append rows and columns, i.e. we pretend they are present.
Another approach is to consider the largest submatrices $A^-$ and $B^-$ of $A$
and $B$ so that the dimensions of $A^-$ and
$B^-$ match our requirements -- this is called ``peeling''. Then once the
product $C^- = A^-B^-$ is computed, one resolves the remaining rows and columns
of $C$ from the remaining rows and columns of $A$ and $B$ that are not in $A^-$
and $B^-$ (cf. \cite{strassen-implementation}). For those remaining pieces
Strassen-Winograd is not used but an implementation which does not cut the
matrices into submatrices. We use the ``peeling'' strategy in our
implementation,
but note that it is easy to construct a case where our strategy is clearly not
optimal, Table~\ref{tab:cutting} gives an example where ``padding'' would only
add one row and one column, while ``peeling'' has to remove many rows and
columns. This is an area for future improvement.

\begin{table}[htbp]
\begin{center}
\begin{tabular}{|c|r|}
\hline
Matrix Dimensions   & Time in $s$ \\
\hline
$2^{14}-1 \times 2^{14}-1$ & 7.86 \\
$2^{14} \times 2^{14}$ & 6.09 \\
$2^{14}+1 \times 2^{14}+1$ & 6.11 \\
\hline
\end{tabular}
\caption{``Peeling'' strategy on 64-bit Linux, 2.33Ghz, \CTD}
\label{tab:cutting}
\end{center}
\end{table}

To represent the submatrices in Algorithm~\ref{alg:strassen} we use ``matrix
windows'' as described earlier. While this has the benefit of
negligible required additional storage compared to out-of-place submatrices,
this affects data locality negatively. To restore data locality, we copy
out the target matrix $C$ when switching from Strassen-Winograd to M4RM. On the
other hand our experiments show that copying out $A$ and $B$ at this crossover
point does not improve performance.
Data locality for $B$ is achieved through the Gray code tables and it appears
that the read of $x$ from $A$ (cf. Algorithm~\ref{alg:m4rm}) does not
significantly contribute to the runtime. 

However, even with ``matrix windows''
Strassen-Winograd requires more memory than classical cubic multiplication.
Additional storage is required to store intermediate results. The most
memory-efficient scheduler (cf. \cite{strassen-memory}) uses two additional
temporary submatrices and is utilized in our implementation. We also tried
the ``proximity schedule'' used in FFLAS~\cite{fflas} but did not see any
improved performance.

\section{Tuning Parameters}
\label{sec:tuning}
Our final implementation calls Strassen-Winograd, which switches over to M4RM
if the input matrix dimensions are less than a certain parameter $c_o$. If $B$
then has fewer columns than $w_s$ (word size in bits)
the classical cubic algorithm is called. This last case is quite common in the
fix-up step of ``peeling''. This strategy gives three parameters for
tuning. The first is $c_o$, the crossover point where we switch from
Strassen-Winograd to M4RM. Second, $b_s$ is the size for block
decomposition inside M4RM for cache friendliness. Third, $k$ dictates the size
of the used Gray code tables. We always fix the number of Gray code tables
to $t=8$.

By default $c_s$ is chosen such that \emph{two} matrices fit into
L2 cache, because this provides the best performance in our experiments. For the
\Opteron (1MB of L2 cache) this results in $c_s=2048$ and for
the \CTD (4MB of L2 cache) this results in $c_s=4096$. We only fit two
matrices, rather
than all three matrices in L2 cache as $b_s$ reduces the size of
the matrices we are working with to actually fit three matrices in L2 cache. The
default value is fixed at $b_s=c_s/2$. The value $k$ is set to
$\lfloor 0.75 \times \log_2 b_s \rfloor - 2$. We subtract 2 as a means to
compensate for the use of 8 Gray code tables. However, if additionally reducing
$k$ by 1
would result in fitting all Gray code tables in L1 cache, we do
that. Thus, $k$ is either $\lfloor 0.75 \times \log_2 b_s \rfloor - 2$ or
$\lfloor 0.75 \times \log_2 b_s \rfloor - 3$ depending on the input dimensions
and the size of the L1 cache.
These values have been determined empirically and seem to provide the
best compromise across platforms.

On the \Opteron these values --- $c_s=2048$, $b_s=1024$,
$k=5$, $t=8$ Gray code tables --- mean that the two input matrices fit into the
1MB of L2 cache, while the 8 Gray code tables fit exactly into L1: $8 \cdot 2^5
\cdot 2048/8 = 64$Kb. The influence of the parameter $b_s$  in the final
implementation is shown in Table~\ref{tab:parameters-opteron} for fixed $k=5$
and $c_s=2048$.

On the \CTD these values are $c_s=4096$, $b_s=2048$, $k=6$, $t=8$ and ensure
that all data fits into L2 cache. Since the \CTD has only 32kb of L1 cache we
do not try to fit all tables into it. So far in our experiments, performance
did not increase when we tried to optimize for L1 cache.

\begin{table}[htbp]
\begin{footnotesize}
\begin{center}
\begin{tabular}{|c|c|c|c|}
\hline
Matrix Dimensions & $b_s=2048$ & {$b_s=1024$} & {$b_s=768$} \\
\hline
$10,000 \times 10,000$ &  2.96  &  2.49 &  2.57\\ 
$16,384 \times 16,384$ & 13.23  & 10.49 & 10.37\\
$20,000 \times 20,000$ & 21.19  & 17.73 & 18.11\\
$32,000 \times 32,000$ & 67.64  & 67.84 & 69.14\\
\hline
\end{tabular}
\caption{Strassen-Winograd multiplication, 64-bit Linux, 2.6Ghz \Opteron}
\label{tab:parameters-opteron}
\end{center}
\end{footnotesize}
\end{table}

\section{Results}
\label{sec:benchmark}
To evaluate the performance of our implementation we provide benchmark
comparisons against the best known implementations we are aware of.
First, \Magma \cite{magma} is widely known for its high performance
implementations of many algorithms. Second, \GAP \cite{GAP4} (or equivalently
the C-MeatAxe \cite{meataxe}) is to our knowledge the best available
open-source implementation of dense matrix multiplication over $\GFZ$. Note,
that the high-performance FFLAS~\cite{fflas}
library does not feature a dedicated implementation for $\GFZ$.
In the Tables~\ref{tab:mult-time-c2d} and~\ref{tab:mult-time-opteron}
we give the average of ten observed runtimes and RAM usage for multiplying two
random square matrices. The timings for M4RI were obtained using
\Sage~\cite{sage}. M4RI was compiled with GCC 4.3.1 on both machines and we used
the options \texttt{-O2} on the \Opteron machine
and \texttt{-O2 -msse2} on the \CTD machine.

\begin{table}[htbp]
\begin{footnotesize}
\begin{center}
\begin{tabular}{|c|r|r|r|r|r|r|}
\hline
  & \multicolumn{2}{c|}{\Magma 2.14-14} &
\multicolumn{2}{c|}{\GAP 4.4.10} & \multicolumn{2}{c|}{M4RI-20080821}\\
\hline
Matrix Dimensions & Time & Memory & Time & Memory & Time & Memory\\
\hline
$10,000 \times 10,000$ &  2.210~s &  85~MB &  6.130~s & 60~MB &  1.504~s & 
60~MB \\
$16,384 \times 16,384$ &  8.670~s & 219~MB & 25.048~s & 156~MB &  6.074~s &
156~MB \\
$20,000 \times 20,000$ & 16.030~s & 331~MB &    --- & --- & 10.721~s & 232~MB\\
$32,000 \times 32,000$ & 58.730~s & 850~MB &    --- & --- & 43.197~s & 589~MB\\
\hline
\end{tabular}
\caption{64-bit Debian/GNU Linux, 2.33Ghz \CTD}
\label{tab:mult-time-c2d}
\end{center}
\end{footnotesize}
\end{table}
%

\begin{table}[htbp]
\begin{footnotesize}
\begin{center}
\begin{tabular}{|c|r|r|r|r|r|r|}
\hline
  & \multicolumn{2}{c|}{\Magma 2.14-13} &
\multicolumn{2}{c|}{\GAP 4.4.10} & \multicolumn{2}{c|}{M4RI-20080811}\\
\hline
Matrix Dimensions & Time & Memory & Time & Memory & Time & Memory\\
\hline
$10,000 \times 10,000$ &  2.656~s &  85~MB &  10.472~s &  60~MB &  2.490~s &
60~MB\\
$16,384 \times 16,384$ & 10.260~s & 219~MB &  43.658~s & 156~MB & 10.490~s &
156~MB \\
$20,000 \times 20,000$ & 18.156~s & 331~MB & --- & --- & 17.730~s & 232~MB \\
$32,000 \times 32,000$ & 67.237~s & 850~MB & --- & --- & 67.840~s & 589~MB \\
\hline
\end{tabular}
\caption{64-bit Debian/GNU Linux, 2.6Ghz \Opteron}
\label{tab:mult-time-opteron}
\end{center}
\end{footnotesize}
\end{table}

\begin{table}[htbp]
\begin{footnotesize}
\begin{center}
\begin{tabular}{|c|r|r|r|r|}
\hline
  & \multicolumn{2}{c|}{\Magma 2.14-16} & \multicolumn{2}{c|}{M4RI-20080909}\\
\hline
Matrix Dimensions & Time & Memory & Time & Memory\\
\hline
$10,000 \times 10,000$ &   7.941~s &  85~MB &   4.200~s &  60~MB\\
$16,384 \times 16,384$ &  31.046~s & 219~MB &  16.430~s & 156~MB\\
$20,000 \times 20,000$ &  55.654~s & 331~MB &  28.830~s & 232~MB\\
$32,000 \times 32,000$ & 209.483~s & 850~MB & 109.414~s & 589~MB\\
\hline
\end{tabular}
\caption{64-bit RHEL 5, 1.6GHz Itanium}
\label{tab:mult-time-itanium}
\end{center}
\end{footnotesize}
\end{table}

\bibliographystyle{plain}
\bibliography{literature}
\label{@lastpg}
\end{document}